\documentclass[%
 reprint,
 amsmath,amssymb,
 aps,
]{revtex4-1}

\usepackage{graphicx}
\usepackage{bm}
\usepackage{amsmath, amssymb}
\usepackage{braket}
\usepackage{float}
\usepackage{chngcntr}

\begin{document}

\preprint{APS/123-QED}

\title{Sensing coherent dynamics of electronic spin clusters in solids}

\author{E. L. Rosenfeld$^{1}$, L. M. Pham$^{2}$, M. D. Lukin$^{1}$, and R. L. Walsworth$^{1,3}$}\email{rwalsworth@cfa.harvard.edu}
\affiliation{%
 $^{1}$Department of Physics, Harvard University, Cambridge, Massachusetts 02138, USA  
 \\
  $^{2}$MIT Lincoln Laboratory, Lexington, Massachusetts 02421, USA
 \\
 $^{3}$Harvard-Smithsonian Center for Astrophysics, Cambridge, Massachusetts 02138, USA 
}%

\date{\today}

\begin{abstract}
We observe coherent spin exchange between identical electronic spins in the solid state, a key step towards full quantum control of electronic spin registers in room temperature solids. In a diamond substrate, a single nitrogen vacancy (NV) center coherently couples to two adjacent $S=1/2$ dark electron spins via the magnetic dipolar interaction. We quantify NV-electron and electron-electron couplings via detailed spectroscopy, with good agreement to a model of strongly interacting spins. The electron-electron coupling enables an observation of coherent flip-flop dynamics between electronic spins in the solid state, which occur conditionally on the state of the NV. Finally, as a demonstration of coherent control, we selectively couple and transfer polarization between the NV and the pair of electron spins. Our observations enable the realization of fast quantum gate operations and quantum state transfer in a scalable, room temperature, quantum processor.
\end{abstract}

\pacs{Valid PACS appear here}
\maketitle
\textit{Introduction.---}Measuring and manipulating coherent dynamics between individual pairs of electronic spins in the solid state opens a host of new possibilities beyond collective phenomena \cite{schweiger, spectral diffusion, silicon,joonhee,lukin mbl}. For example, a quantum register consisting of several coherently coupled electronic spins could serve as the basic building block of quantum information processors and quantum networks \cite{wrachtrup qm register, paola nuc spin register, wrachtrup nuc spin register}. Additionally, recent proposals indicate that dynamics between many unpolarized electronic spins can mediate fully coherent coupling between distant qubits to be used for quantum state transfer \cite{norm, ashok qst, gualdi, paola NMR}; measuring the coherent flip-flop rate between a pair of electronic spins could allow for sensitive distance measurements in individual molecules in nanoscale magnetic resonance imaging \cite{blank spin diffusion,blank flip flop}. 

However, such an interaction is a challenge to observe \cite{blank flip flop, blank spin diffusion}. In particular, the identical spins need to be close enough to interact strongly, such that the spins cannot be spatially or spectrally resolved, to allow for polarization exchange. In prior work, polarization transfer was measured between either spatially or spectrally resolved electronic spins: e.g., between two Strontium-88 ions separated by $\mu$m scales \cite{nature 2014 ions} or between a nitrogen vacancy (NV) color center and a substitutional nitrogen in diamond \cite{wrachtrup p1, awschalom p1, helena, chinmay}. Conversely, nuclear spin-spin dynamics have been observed in diamond, facilitated by long nuclear spin coherence times and using a single NV center as a mediator \cite{lily, gurudev,rep readout}. Control of NV-nuclear spin clusters has led to using nuclear spins as a room temperature quantum memory and quantum register \cite{rep readout, taminau, kalb natcomm, abobeih arxiv}, with applications such as NMR detection of a single protein \cite{igor} and quantum networks \cite{delft,taminau}. Similarly, manipulating interactions between identical electronic spins could lead to faster gate times and long-distance transport in solid state, room-temperature quantum information processors \cite{norm}, features that are challenging for nuclear spins due to their weaker coupling strengths.

\begin{figure}[h!]
\centering
\includegraphics[scale=0.75]{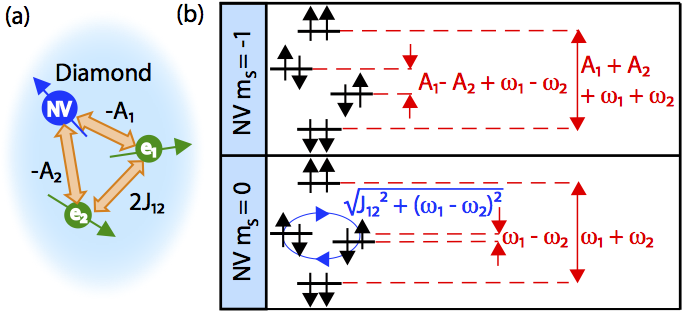}
\caption{\label{fig:fig1} (a) Schematic of a three-electronic-spin cluster in diamond, labeled with coupling strengths. (b) Energy level diagram of two dipolar-coupled dark electron spins (each S = 1/2) as a function of the nearby NV spin state (S = 1). When the NV is in the $\ket{-1}_{\text{NV}}$ spin state, the magnetic field gradient it produces at the electrons suppresses their dynamics. When the NV is in the $\ket{0}_{\text{NV}}$ spin state, flip-flops are allowed between the electron spins.}
\end{figure}

Here, we report coherent spin exchange between two identical electronic spins, a vital prerequisite for many of the ideas discussed above, including the aforementioned collective phenomena \cite{schweiger, spectral diffusion, silicon,joonhee,lukin mbl}. A single NV center acts as a nanoscale probe of flip-flop interactions between a pair of electron spins. First, we identify a coherently-coupled, three-spin cluster consisting of the optically-active NV and two optically-dark electron spins inside the diamond [Fig. \ref{fig:fig1}(a)]. The coupling strengths and resonance frequencies for the three spins are extracted via optically detected magnetic resonance (ODMR) NV spectroscopy, as well as dynamical decoupling and double electron-electron resonance (DEER) experiments. The electron spins undergo flip-flop dynamics, conditional on the state of the NV [Fig. \ref{fig:fig1}(b)], as in a controlled SWAP gate. Finally, we demonstrate partial manipulation of the three-electronic-spin cluster through selective coupling and transfer of polarization between the NV and the pair of electron spins. 
\begin{figure}[h!]
\centering
\includegraphics[scale = 0.63]{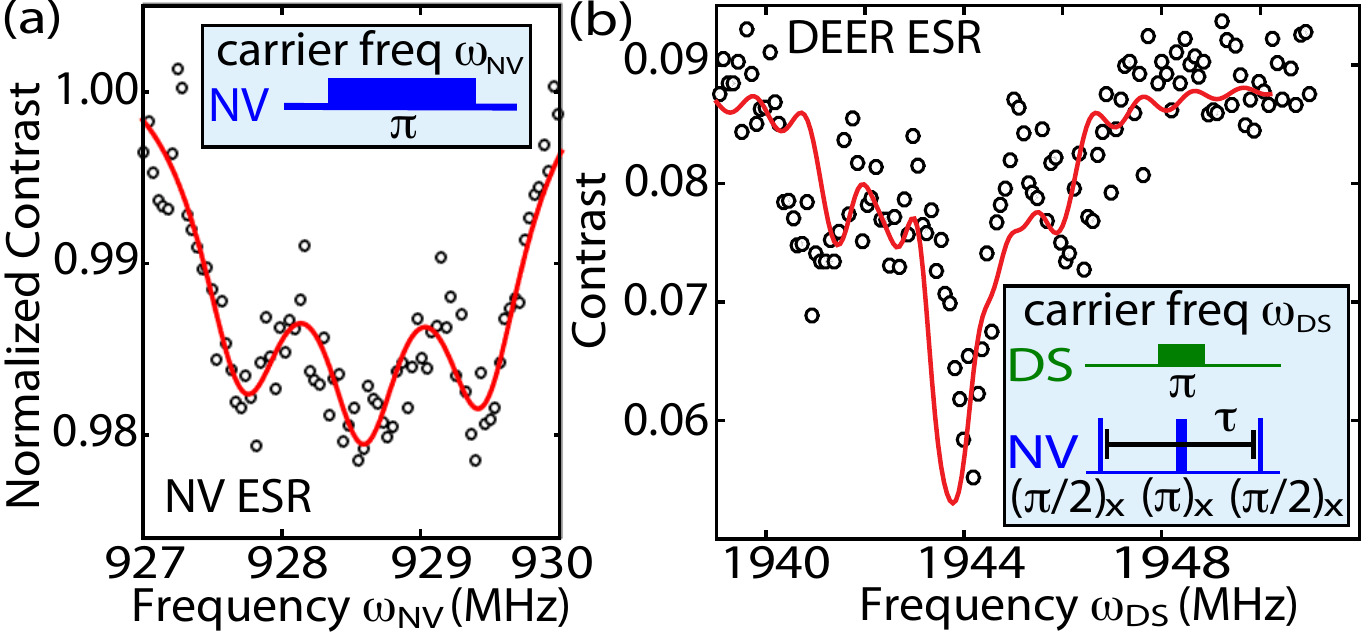}
\caption{\label{fig:fig2} Spectroscopy of three electronic-spin cluster. (a) Measured NV ESR spectrum (black circles), in the presence of a $B_0$ = 694.0(6) G static bias field, fit to three Lorentzian curves (red line). The triplet-like structure is consistent with the model and parameter values presented in this work. The $^{14}$N nuclear spin is polarized into the $m_I=+1$ state due to the large transverse NV-$^{14}$N hyperfine coupling in the optically excited manifold \cite{ran fisher, v jacques}. The estimate of the $B_0$ field is adjusted accordingly. Inset, NV ESR pulse sequence. (b) DEER ESR spectrum data (black dots), in the presence of a $B_0$ = 694.0(6) G bias field, with numerical simulation from equation (\ref{eqn1}) using parameter values given in the main text (red line). Time $\tau$ is fixed to 3 $\mu$s. Observed lineshape is qualitatively consistent with two electrons strongly coupled to both each other and the probe NV. Inset, DEER ESR pulse sequence.}
\end{figure}

\textit{Experimental results.---}The unpolished diamond sample features a 99.999\% $^{12}$C epitaxially grown layer, implanted with $^{14}$N ions at 2.5 keV and annealed for eight hours at 900 $^\circ$C. A mask implantation was performed, such that the density of implanted nitrogen varied from close to zero to $10^{12}$/cm$^2$ across the sample. Measurements were performed using a custom-built confocal microscope with a 532 nm laser for NV excitation, and a single photon counter to collect phonon sideband photoluminescence for population readout of the NV ground state sublevels. A dual-channel arbitrary waveform generator enables coherent driving of the NV spin and two additional electron spins in the diamond. The NV and electron spin levels are split by a DC magnetic field ($B_0=694.0(6)$ G) aligned along the NV axis and generated by a permanent magnet. 

An electron spin resonance (ESR) measurement on the NV reveals an atypical spectrum. Figure \ref{fig:fig2}(a) illustrates the atypical ESR spectrum containing a triplet-like structure, with splitting about a factor of 2.5 smaller than the $^{14}$N hyperfine coupling \cite{newton nv hyp}. Fitting the data to three Lorentzian lineshapes demonstrates a full splitting of 1.70(7) MHz. 

To determine if this characteristic splitting is explained by the presence of spins with electronic character, we selectively drive the spins with resonances around $\gamma_eB_0$, using a separate microwave channel (labeled DS for ``dark spin" in Figure \ref{fig:fig2}(b), inset). When the DS drive frequency approaches a resonance of an electron spin coupled to the NV, the NV Bloch vector accumulates phase in the transverse plane as in a Double-Electron-Electron-Resonance spectroscopy (DEER ESR) experiment. With a central dip around $g=2$, the spectrum shows a characteristic, asymmetric lineshape [Fig. \ref{fig:fig2}(b)], for which either nuclear quadrupolar spin(s) strongly coupled to a single electron, or dipolar coupling(s) between multiple electronic spins could be responsible. 

\begin{figure*}[t!]
\centering
\includegraphics[width = \textwidth]{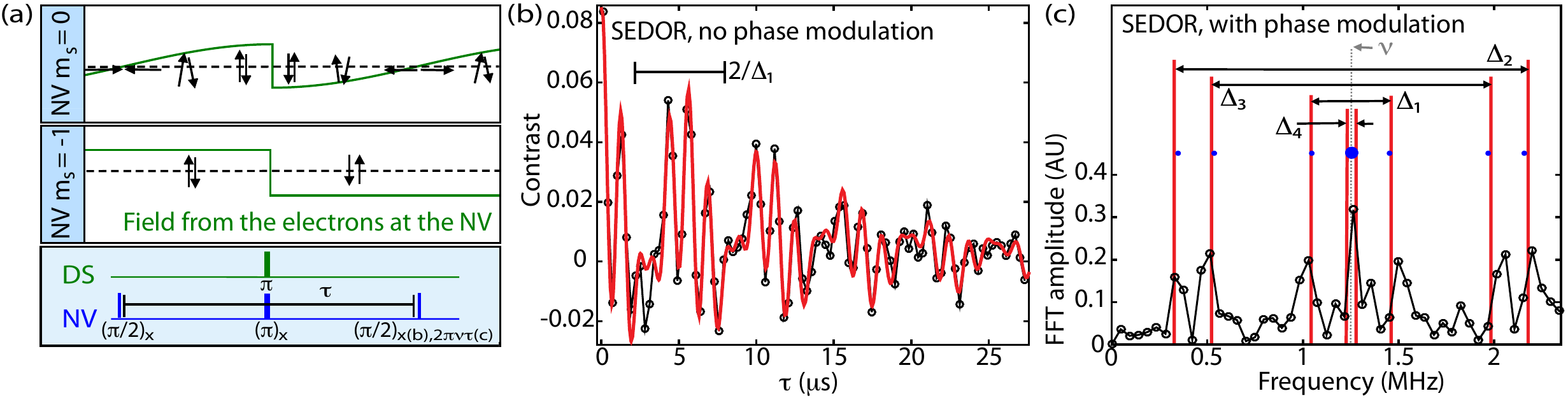}
\caption{\label{fig:fig3} Coherent dynamics of the three-electronic-spin cluster. (a) Lowest panel: SEDOR pulse sequence schematic. Uppermost panel: depiction of the electron dynamics corresponding to NV population in $\ket{0}_{\text{NV}}$. Middle panel: electron spin evolution corresponding to $\ket{-1}_{\text{NV}}$. (b) Time-domain data (black circles and line) of the SEDOR experiment. The solid red line is a fit to four sine waves multiplied by a decaying exponential (T$_2$ = 14(3) $\mu s$, p = 1.1(4)). The frequencies from the fit are consistent with the frequencies reported with the model and parameter values \cite{suppl}. The period $2/\Delta_1$ corresponding to the electron flip-flop dynamics is shown. Data was taken at 180 G. (c) Fourier transform of data from the SEDOR experiment performed with phase modulation (TPPI) on the last NV $\pi/2$ pulse at a frequency $\nu=1.25\text{ MHz}$ (black dots and line). The signal amplitudes for each frequency pair about $\nu$ are equal, consistent with two unpolarized electron spins. Blue dots correspond to the frequencies found in the fit in (b), up-converted by the TPPI frequency $\nu$. Error bars (95\% CI) from the fit are a factor of four smaller than the diameter of the dots, except for the large dot at $\nu$, for which the error is the size of the dot. The vertical red lines represent frequency components $\Delta_{1-4}$ corresponding to the analytical solution using the model and parameter values reported in the main text \cite{suppl}.}
\end{figure*}

Distinguishing between these possibilities requires a study of the number of electronic spins present. In a Spin Echo DOuble Resonance (SEDOR) pulse sequence \cite{schweiger} [Fig. \ref{fig:fig3}(a), bottom panel], a single electron spin induces oscillations in the NV population, and hence the ODMR signal, at the frequency of the NV-electron dipolar coupling strength. However, the presence of multiple electronic spins results in multiple frequencies, originating from the different coupling strengths (electron-electron, NV-electron), as well as any coherent dynamics. The resulting data exhibits several frequency components [Fig. \ref{fig:fig3}(b)], consistent with a coherently-coupled, multi-electronic spin system [Fig. \ref{fig:fig1}(a)]. Comparing the observed Rabi frequencies of the NV and electronic spin transitions confirms that the dark electron spins are S = 1/2 \cite{suppl}.

\textit{Model and Hamiltonian.---}The triplet lineshape components extracted from the NV ESR, as well as the frequencies in the SEDOR measurement [Fig. \ref{fig:fig3}(b)], are well-described by a system of two electron spins coherently coupled to the NV. The three-spin cluster is modeled using the following Hamiltonian, in the secular approximation and frame rotating at the NV transition frequency:
\begin{multline}
\frac{\mathcal{H}}{h} = \sum_{i = 1,2} {\Big(\omega_i + A_i(S_{z}^{\text{NV}} + {\bf{I}} /2)\Big)}S_{z}^{(i)} +\\ 
J_{12} \Big(2 S_{z}^{(1)}S_{z}^{(2)} - \frac{1}{2}(S_{+}^{(1)} S_{-}^{(2)} + S_{-}^{(1)} S_+^{(2)})\Big).
\label{eqn1}
\end{multline}Here, $-A_ih=-\mu_0\hbar^2\gamma_e^2(3\cos^2⁡{\theta_i}-1)/(4\pi r_i^3)$ is the magnetic dipole interaction strength between the NV and electron $i$. The electron-electron coupling term $J_{12}h=-\mu_0\hbar^2\gamma_e^2(3\cos⁡^2{\theta_{12}}-1)/(8\pi r_{12}^3)$ is half the magnetic dipole interaction strength between the electrons, and $\omega_{i}$ is the Zeeman energies of electron $i$. Note that the $\ket{m_s=+1}\equiv\ket{+1}_{\text{NV}}$ state is not populated under the experimental conditions employed in this work, reducing the NV subspace to $\ket{0}_{\text{NV}}$ and $\ket{-1}_{\text{NV}}$ in equation (\ref{eqn1}). Therefore, all of the operators are 2x2 spin matrices.

An analytical calculation of the SEDOR signal using the Hamiltonian in equation \ref{eqn1} yields four characteristic frequencies (labeled $\Delta_{1-4}$), which are functions of $J_{12}$, $A_1$, $A_2$, and $\omega_1-\omega_2$ \cite{suppl}. We find good agreement between the SEDOR data and a sum of four sine waves (one of which is below the spectral resolution of the current experiment), multiplied by $e^{-(t/T_2)^p}$ to account for NV decoherence [Fig. \ref{fig:fig3}(b)]. To extract the parameter values, we associate the three resolved frequencies with the predicted frequency-domain behavior from the model \cite{suppl} and solve for $J_{12}$, $\omega_1-\omega_2$ and $A_1-A_2$, obtaining an upper bound on $A_1+A_2$ from the unresolved frequency component. We impose agreement with the observed NV ESR and DEER ESR spectrum to confirm our solution and inform the value of $\omega_1$, as well as the value of $A_1+A_2$ \cite{suppl}. The resulting parameter values reported here are $A_{1,2}=0.81(5),-0.86(5)\text{ MHz}$; $J_{12}=\pm0.38(5)\text{ MHz}$; $\omega_1=\gamma_eB_0+0(2)\text{ MHz}$ and $\omega_2=\omega_1-0.14(5)\text{ MHz}$ \cite{suppl}. 

As mentioned above, this model is also consistent with the observed NV ESR and DEER ESR spectra [Fig. \ref{fig:fig2}]. Two of the eigenstates of the electron pair, $\ket{\uparrow\downarrow}$, $\ket{\downarrow\uparrow}$, each induce a dipolar magnetic field of strength $\pm(A_{1}-A_{2})/2=\pm0.84(4)\text{ MHz}$, which consequently splits the NV ESR lines. The two other electron pair states, $\ket{\downarrow\downarrow}$ and $\ket{\uparrow\uparrow}$, exert a field with strength $\pm(A_{1}+A_{2})/2=\pm0.03(4)\text{ MHz}$. The result is an NV triplet-like spectrum, with splittings given by the difference of the NV-electron couplings. As mentioned above, fitting the NV ESR data [Fig. \ref{fig:fig2}(a), black dots] to a sum of three Lorentzian curves confirms the NV resonance frequencies, which are split by 1.70(7) MHz (95\% CI of the fit), in good agreement with the model parameters $A_{1}-A_{2}=1.67(7)\text{ MHz}$.

Conversely, the presence of multiple electrons corrupts the direct measurement of individual transition frequencies in the DEER ESR spectrum. The DEER ESR lineshape depends sensitively on all coupling and resonance frequency parameters \cite{suppl}, which we calculate numerically with the model. Using the same parameter values listed above, we demonstrate good qualitative agreement between the DEER ESR data [Fig. \ref{fig:fig2}(b), black dots] and the model [Fig. \ref{fig:fig2}(b), red line], within the error ranges on the extracted model parameters \cite{suppl}.

\begin{figure}[h!]
\centering
\includegraphics[scale = 0.85]{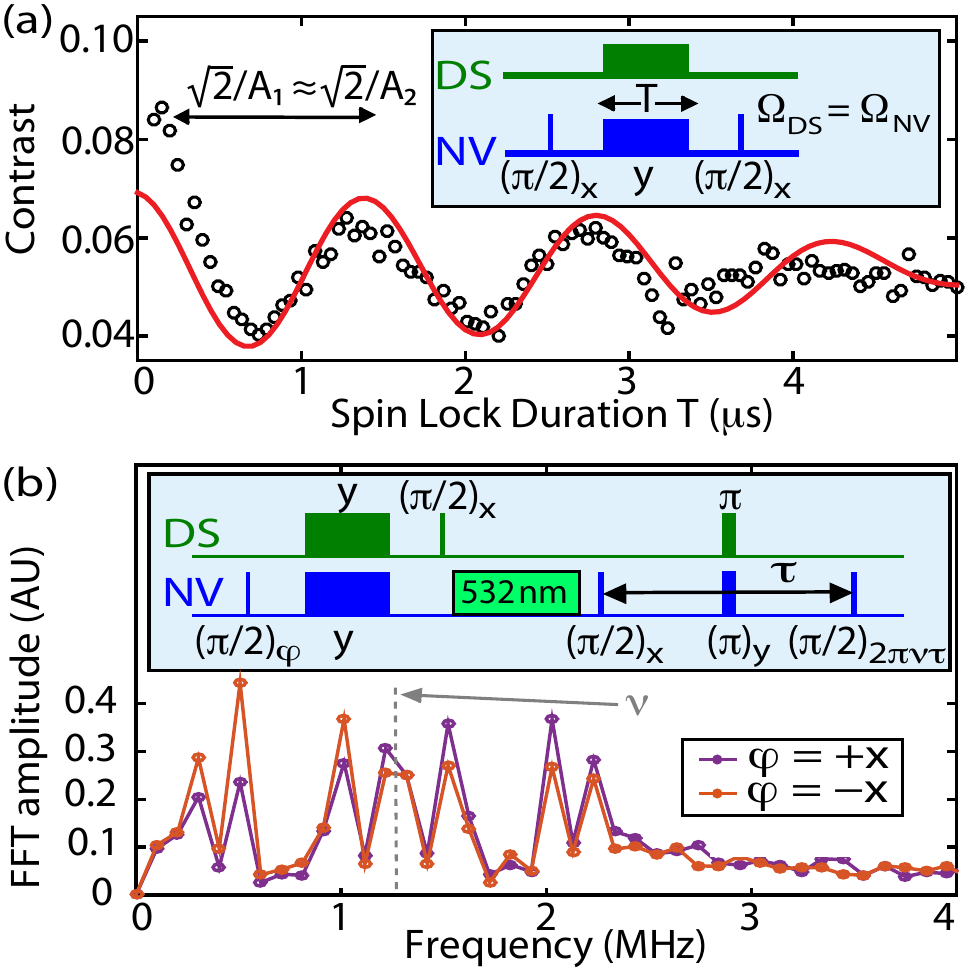}
\caption{\label{fig:fig4} Coherent polarization transfer between the NV and two dark electron spins. (a) Observed polarization transfer to the electrons at the Hartmann-Hahn resonance condition \cite{hartmann hahn} as a function of spin lock duration T (black circles). Red line is a numerical simulation of the experiment protocol using equation (\ref{eqn1}) and the parameter values given in the main text. The primary oscillation frequency of 0.60(5) MHz, $\simeq A_1/\sqrt{2}\simeq A_2/\sqrt{2}$, is consistent with the two similar NV-electron coupling strengths adding in quadrature. Due to microwave amplitude instability of the setup, we allow for a detuning from the resonance condition up to 350 kHz. Effects not included in our model that contribute to the deviation at short T are: slow drifts in this detuning, dephasing of detuned driving of (weakly populated) hyperfine transitions, and pulse errors of the initial NV $\pi/2$ pulse. Inset, Hartmann-Hahn pulse sequence. (b) Observed polarization transfer to the electrons using Hartmann-Hahn cross polarization at fixed spin lock duration T = 700 ns, followed by optical repolarization of the NV to $\ket{0}_{\text{NV}}$, then readout of the electron pair polarization via SEDOR. Results are displayed as the FFT of the SEDOR data. The phase $\phi$ of the first NV pulse determines the direction of polarization transfer. For both directions, the polarization is shared across the three frequency pairs, consistent with two coupled electron spins. Inset, experimental pulse sequence.}
\end{figure} 

\textit{Coherent dynamics in the cluster.---}An understanding of the three-electronic-spin cluster allows for a discussion of the coherent dynamics between the electron spins. When the $\ket{-1}_{\text{NV}}$ spin state is occupied, two of the electron-pair energy levels, $\ket{\uparrow\downarrow}$ and $\ket{\downarrow\uparrow}$, differ by $A_{1}-A_{2}+\omega_{1}-\omega_{2}=1.81(9)\text{ MHz}$, which is larger than their coupling strength $J_{12}=\pm0.38(5)\text{ MHz}$; thus, flip-flops are suppressed [Fig. \ref{fig:fig1}(c), top panel]. However, when the NV population occupies the $\ket{0}_{\text{NV}}$ spin state, the same two energy levels are split by only $\omega_{1}-\omega_2=0.14(5)\text{ MHz}$, allowing for polarization exchange [Fig. \ref{fig:fig1}(c), bottom panel]. Direct diagonalization of the Hamiltonian shows that flip-flops occur at rate $\Delta_1\equiv\sqrt{J_{12}^2+(\omega_1-\omega_2)^2}=0.41(5)\text{ MHz}$ [Fig. \ref{fig:fig1}(b), bottom panel].

In the SEDOR pulse sequence, sweeping the free precession time $\tau$ and fixing the electron spin $\pi$-pulse on resonance allows for quantitative observations of the flip-flop frequency between the electrons. During the time $\tau$, the NV Bloch vector accumulates phase in the transverse plane due to the dipolar field of the electrons, described by the $S_{z}^{\text{NV}}S_{z}^{(i)}$ terms in equation (\ref{eqn1}). Since half of the NV population is in the $\ket{0}_{\text{NV}}$ spin state throughout this measurement, dynamics between the pair of electrons are partially allowed [Fig. \ref{fig:fig3}(a), top and middle panels]. Sweeping the free precession time constitutes an AC magnetometer, where the AC field amplitude of 0.84(4) MHz is generated by the pair of electrons in the $\ket{\uparrow\downarrow}$ or $\ket{\downarrow\uparrow}$ states. The detected AC field frequency $\Delta_{1}/2$ is given by half the electron spin pair flip-flop rate [Fig. \ref{fig:fig3}(a), top panel], and is marked in the time domain in Figure \ref{fig:fig3}(b). As constructed, the frequency components of the SEDOR data implies $\Delta_1=0.41(5)\text{ MHz}$, equal to the value found with the model parameters \cite{suppl}.

In addition, the parameters $A_{1}$ and $A_{2}$ contribute to other frequency components. During the SEDOR sequence, the other half of the NV population occupies the $\ket{-1}_{\text{NV}}$ spin state, such that the electron-pair dynamics are suppressed by the field of strength $A_{1} - A_{2}=1.67(7)\text{ MHz}$ [Fig. \ref{fig:fig3}(a), middle panel]. The eigenstates of the relevant Hamiltonian $H_{-1}^{\text{DS}}$ are mostly described by the Zeeman $\ket{\uparrow\downarrow}$ and $\ket{\downarrow\uparrow}$ states, and are dressed by their interaction, which shifts their energy splitting. These states consequently modulate the SEDOR data at rate $\frac{1}{2}\sqrt{(A_1-A_2+\omega_1-\omega_2)^2+J_{12}^2}\equiv\Delta_{2}/2=0.93(4)\text{ MHz}$, equal to a frequency component observed in the SEDOR data \cite{suppl}.

Throughout the pulse sequence, the NV is in a coherent superposition of the $\ket{0}_{\text{NV}}$ and $\ket{-1}_{\text{NV}}$ spin states. The resulting interference of both electron propagators induces additional frequency components in the NV evolution at half the sums and differences of $\Delta_{1}$ and $\Delta_{2}$ \cite{suppl}. We find that the amplitude of the $(\Delta_{1}+\Delta_{2})/2$ frequency component decreases via destructive interference of the two propagator paths, due to the relative detuning between the two electrons $\omega_1-\omega_2$ \cite{suppl}; similarly, the amplitude of the $(\Delta_{2}-\Delta_{1})/2\equiv\Delta_3/2=\frac{1}{2}(\sqrt{(A_1-A_2+\omega_1-\omega_2)^2+J_{12}^2}-\sqrt{J_{12}^2+(\omega_{1}-\omega_{2})^2})=0.72(2) \text{ MHz}$ component increases \cite{suppl}. As expected, the fit to the SEDOR data also exhibits a frequency component at 0.72(2) MHz \cite{suppl}. Finally, irrespective of the state of the NV, the states $\ket{\uparrow\uparrow}$ and $\ket{\downarrow\downarrow}$ modulate the SEDOR signal at the frequency $\Delta_{4}/2\equiv(A_1+A_2)/2$. For the present three-spin system, we estimate $\Delta_4/2=-0.03(4)\text{ MHz}$, which is not distinguishable from zero for the present experiment.

As a check of reproducibility, we repeat the SEDOR experiment using a phase modulation technique \cite{suppl}, known as time-proportional phase increments (TPPI) in nuclear magnetic resonance, to up-convert the signals away from zero frequency by $\nu=1.25\text{ MHz}$ \cite{schweiger} [Fig. \ref{fig:fig3}(c)]. The Fourier transform of the TPPI data [Fig. \ref{fig:fig3}(c), black line] shows pairs of spectral peaks at frequencies corresponding to $\Delta_{1-3}$, centered around the TPPI frequency $\nu$; as before, the $\Delta_4$ peaks are not resolved. The positive and negative frequency components of each pair have approximately equal amplitude, consistent with unpolarized electron spins. The red lines corresponding to $\Delta_{1-4}$ in Figure \ref{fig:fig3}(c) indicate the expected frequencies from model. The frequency components from the fit of the time domain data [Fig. \ref{fig:fig3}(b)] are up-converted by $\nu$ and marked as blue dots, and agree with the model within the margin of error \cite{suppl}.

\textit{Manipulation of the electronic spins.---}Finally, we demonstrate coherent manipulation of the three-spin cluster by transferring polarization from the NV to the dark electron spin pair using a Hartmann-Hahn technique \cite{hartmann hahn}. We first fix the amplitude of the drives to the Hartmann-Hahn resonance condition \cite{hartmann hahn}, and transfer polarization from the NV to the electron spins while sweeping the spin lock duration T [Fig. \ref{fig:fig4}(a)]. By matching the dressed state energies of the NV and dark spins, NV-dark spin flip-flops become allowed and the dark spins are polarized. By energy conservation, the dark spins are aligned parallel (anti-parallel) to the resonant drive vector in the rotating frame, if the NV Bloch vector is initialized parallel (anti-parallel) along the NV drive vector. The polarization evolves from the NV and returns at a rate approximately given by $A_1/\sqrt{2}\approx A_2/\sqrt{2}$, as expected for two uncorrelated electrons with approximately equal coupling to the NV. Next, we observe polarization of the dark electron spin pair by fixing the spin lock duration at T = 700 ns $\approx 1/|\sqrt{2}A_1| \approx 1/|\sqrt{2}A_2|$, re-polarizing the NV with a 532 nm laser pulse, and reading out the polarization of the electron spins using SEDOR and TPPI. Changing the phase of the first $\pi/2$ pulse on the NV, and therefore the initial NV dressed state, exchanges the direction of polarization transfer [Fig. \ref{fig:fig4}(b), orange and purple lines]. Adding a $\pi/2$ pulse on the dark spins after the spin lock pulse stores the dark spin polarization along the quantization axis. For both polarization transfer directions, the difference in peak amplitude is spread across all pairs of frequencies $\Delta_{1-3}$ [Fig. \ref{fig:fig4}(b)], as is expected for a coupled pair of electrons. Compared to previous work \cite{chinmay,helena}, this constitutes a measurement of coherent polarization transfer from the NV to electron spins, followed by readout of the polarization, opening the door to quantitative estimates of dark spin state preparation fidelities. Here, a careful study of the polarization fidelity will require stringent microwave amplitude stability during a two-dimensional sweep of T and $\tau$, beyond the scope of this work. 

\textit{Outlook.---} Our observations of coherent dynamics between nearby electronic spins in the solid state, under ambient conditions and without spectrally or spatially resolved spins, constitutes a key step toward realizing coherent quantum manipulation of electronic spins. Specifically, the demonstrated techniques can be used to implement quantum registers with fast gate time and quantum state transfer between remote spins via an intermediate spin bath \cite{norm,ashok qst,paola NMR,gualdi}. Additionally, electronic spin dynamics external to an NV could enable a range of potential sensing applications. For example, it can be employed following a recent proposal to measure the spin diffusion rate between intra-molecular spin labels in biomolecules \cite{blank spin diffusion,blank flip flop}, to obtain improved distance measurements beyond the standard DEER protocol \cite{DEER, DEER2}. 

We are indebted to R. Landig, S. Choi, D. Bucher, A. Sushkov, H. Knowles, J. Gieseler, P. Cappellaro, T. van der Sar, E. Bauch, C. Hart, M. Newton and B. Green for fruitful discussions, and especially A. Cooper for informing the authors about the phase modulation technique. The authors are grateful to J. Lee for fabricating the microwave stripline and A. Ajoy for implanting the diamond sample. This material is based upon work supported by the National Science Foundation Graduate Research Fellowship Program under Grant No. DGE1144152 and DGE1745303. Any opinions, findings, and conclusions or recommendations expressed in this material are those of the author(s) and do not necessarily reflect the views of the National Science Foundation. This work was performed in part at the Center for Nanoscale Systems (CNS), a member of the National Nanotechnology Coordinated Infrastructure Network (NNCI), which is supported by the National Science Foundation under NSF award no. 1541959. CNS is part of Harvard University. This material is based upon work supported by, or in part by, the U. S. Army Research Laboratory and the U. S. Army Research Office under contract/grant numbers W911NF1510548 and W911NF1110400. This work was additionally supported by the NSF, Center for Ultracold Atoms (CUA), Vannever Bush Faculty Fellowship and Moore Foundation.

\pagebreak
\pagebreak 
\widetext
\begin{center}
\textbf{\large Supplementary Material to ``Sensing coherent dynamics of electronic spin clusters in solids''}
\end{center}

\setcounter{equation}{0}
\setcounter{figure}{0}
\setcounter{table}{0}
\counterwithin{figure}{section}
\counterwithin{table}{section}
\section{Experimental setup}
\subsection{Optical setup}
We use a home-built 4f confocal microscope to initialize and read out the NV photoluminescence, and a 532 nm green diode laser (Changchun New Industries Optoelectronics Tech Co, Ltd, MGL H532), for NV illumination and initialization. The laser pulses are modulated using an acousto-optic modulator (IntraAction ATM series 125B1), which is gated using a pulseblaster card (PulseBlasterESR-Pro, SpinCore Technologies, Inc, 300 MHz). We use an initialization pulse duration of 3.8 $\mu$s, of which $880$ ns is used for readout of the NV ground state. The NV fluorescence is filtered using a dichroic mirror (Semrock FF560-FDi01) and notch filter (Thorlabs FL-532), and read out using a single photon counter (Excelitas Technologies, SPCM-AQRH-13-FC 17910). A galvonometer (Thorlabs GVS002) scans the laser in the transverse plane, and a piezoelectric scanner placed underneath the objective (oil immersion, Nikon N100X-PFO, NA of 1.3) is used to focus. The diamond is mounted on a glass coverslip patterned with a stripline for microwave delivery (see section 1.2), which is placed directly above the oil immersion objective lens.
\subsection{Microwave driving of the NV and electron spins}
A dual-channel arbitrary waveform generator (Tektronix AWG7102) is used to synthesize the waveforms for the NV and electron spin drives for all pulse sequences. The AWG output is triggered using the pulseblaster card. The NV drive and electron spin drives are amplified  separately (Minicircuits ZHL-42W+, Minicircuits ZHL-16W-43-S+, respectively), then combined using a high power combiner (Minicircuits ZACS242-100W+). The microwave signals are then connected to an $\Omega$-shaped stripline, 100 $\mu$m inner diameter, via a printed circuit board. The diamond is placed onto the center $\Omega$ of the stripline. Typical Rabi frequencies for the NV and electron spins (around 900 and 2000 MHz, respectively) are between 10-15 MHz. 

\subsection{Diamond sample}
The diamond substrate was grown by chemical vapor deposition at Element Six, LTD. It is a polycrystalline, electronic grade substrate. A 99.999\% 12-C layer was grown on the substrate, also at Element Six, along the \{110\} direction, and the diamond was left unpolished. Implantation was performed at Ruhr Universität in Bochum, Germany. Both atomic and molecular ions, $^{14}$N and $^{14}$N$_{2}$, were implanted separately, in a mask pattern with confined circles approximately 30 $\mu$m in diameter, at 2.5 keV implantation energy. The density of implantation varies from 1.4x10$^{12}$/cm$^2$ to 1.4x10$^{9}$/cm$^2$ in different mask regions. For the present study, an NV found near the 1.4x10$^{12}$/cm$^2$ ion implant region is used. The sample was annealed in vacuum at 900 degrees C for 8 hours. 

About 400-500 NVs were investigated, and over 75\% were found to have poor contrast or the incorrect orientation in the bias field. Of the total number of NVs investigated, 113 NVs were screened for coupling to dark spins. The NVs are implanted to be about 5-10 nm below the surface, such that there is a high probability to detect surface dark spins (we expect from e.g. dangling bonds close to the diamond edge). Of the 113 NVs screened for dark spin coupling, about 9 demonstrated coherent coupling to dark spin(s), 63 demonstrated incoherent coupling to a bath of dark spins, and 41 exhibited no signature of dark spin coupling. The individual NV described in our manuscript had coherent coupling to multiple dark spins, and was also very stable in its behavior (for more than a year while the experiments were performed).  A frontier challenge in the NV-diamond community is to fabricate, reliably and predictably, NVs and dark spins with coherent couplings and other optimal properties, so that such mass screenings are not necessary.
\subsection{Static bias field}
We use a permanent magnet (K\&J Magnetics, Inc, DX08BR-N52) to induce a magnetic field \textit{B}$_0$ and thereby split the NV and electron spin energy levels. The diamond is mounted on a three-axis translation stage, which is controlled by three motorized actuators (Thorlabs Z812B). To align the magnetic field to the NV axis, we sweep the position of the magnet, monitor the NV fluorescence (which decreases with field misalignment due to mixing of the magnetic sublevels) and fix the magnet position to the point of maximum count rate. Field alignment precision is about 2 degrees, given by the shot noise of our fluorescence measurements. In order to stabilize the magnitude of the \textit{B}$_0$ field at our system during an experiment, we periodically (approximately every 15 minutes) measure the $m_s = 0 \rightarrow -1$ transition frequency of the NV, and move the magnet position to stabilize the NV transition frequency, and therefore field, at a particular value. Using this technique, we can stabilize the field enough to resolve the $\sim$1 MHz splittings in the DEER ESR experiment, despite the 0.1\%/K temperature coefficient of Neodymium magnets \cite{tempco}. Our \textit{B}$_0$ measurement is taken from the NV transition frequency, which therefore includes error from the 2 degree field misalignment, as well as the approximate error of a measurement of this particular NV zero field splitting, of 2.872(2) GHz, to give a \textit{B}$_0$ measurement accuracy of about 0.1\%.

\section{Signal frequencies and amplitudes in the SEDOR experiment}

In this section we discuss the frequencies seen in the SEDOR experiment, referring to Figs. 3(b) and 3(c) in the main text. 

As stated in the main text, the Hamiltonian is: 
\begin{align}
\frac{\mathcal{H}}{h} = \sum_{i = 1,2} {\Big(\omega_i + A_i\Big(S_{z}^{\text{NV}} + {\bf{I}} /2\Big)\Big)}S_{z}^{(i)} + 
\\J_{12} \Big(2 S_{z}^{(1)}S_{z}^{(2)} - \frac{1}{2}(S_{+}^{(1)} S_{-}^{(2)} + S_{-}^{(1)} S_+^{(2)})\Big)
\end{align}

Where $\omega_1 = \gamma_e B_{0} + 0(2) \text{ MHz}$, $\omega_2 = \omega_1 - 0.14(5) \text{ MHz}$, are the resonance frequencies electron spins 1 and 2, $A_{1,2}=0.81(5),-0.86(5) \text{ MHz}$, are the negative of their couplings to the NV, and $J_{12}= \pm 0.38(5) \text{ MHz}$ is half the electron spin -- spin coupling. 

The states of the electron spins $\ket{\uparrow\uparrow}$ and $\ket{\downarrow\downarrow}$ are decoupled from the states $\ket{\uparrow\downarrow}$ and $\ket{\downarrow\uparrow}$, because the Hamiltonian and SEDOR pulse sequence conserve $|S_{z}^{(1)} + S_{z}^{(2)}|$. Therefore, we consider those two subspaces separately.

For the electron spin states $\ket{\uparrow\uparrow}$ and $\ket{\downarrow\downarrow}$ the flip-flop terms are zero, and the Ising interaction term $S_{z}^{(1)}S_{z}^{(2)}$ is a constant, so the only relevant terms are: 

\begin{align}
\frac{\mathcal{H}}{h} = \sum_{i = 1,2} {\Big(\omega_i + A_i\Big(S_{z}^{\text{NV}} + {\bf{I}} /2\Big)}\Big)S_{z}^{(i)}.
\end{align}

The NV accumulates phase due to its the secular dipolar coupling to the electron spins. Therefore, the NV signal rotates at $\Delta_{4}/2 = (A_1 + A_2)/2$ before phase modulation. 

For the $\ket{\uparrow\downarrow}$ and $\ket{\downarrow\uparrow}$ states, we need to consider the dynamics. Treating this subspace as a two level system where $\ket{\Uparrow} = \ket{\uparrow\downarrow}$ and $\ket{\Downarrow} = \ket{\downarrow \uparrow}$ and dropping the Ising interaction term, we have: 

\begin{align}
\frac{\mathcal{H}}{h} = \Big(\delta + a(S_{z}^{\text{NV}} + {\bf{I}} /2)\Big)s_{z} + J_{12} s_{x}.
\end{align}

Where $\delta = \omega_{1}-\omega_{2}$ and $a = A_{1}-A_{2}$. We use $s_{i}$ to represent the new S = 1/2 spin operators. Here we define two Hamiltonians: one for the NV population in $m_s = 0$, defined as $\mathcal{H}_{0}/h = \delta s_{z} + J_{12} s_{x}$, and conversely one for the NV population in $m_s = -1$, defined as $\mathcal{H}_{-1}/h = (\delta + a)s_{z} + J_{12} s_{x}$. We use the operators $s_{x,y,z}$ to denote the new electron spin operators in the new 2x2 subspace. Let the unitaries $U_{0, -1}(t) \equiv e^{-i \mathcal{H}_{0, -1}t/\hbar}$. 

At the start of the pulse sequence, with the NV starting in $\ket{-1}_{\text{NV}}$, and immediately following the first $\pi/2$ pulse on the NV about y, the density matrix is:

\begin{align}
\rho(0) = (\ket{0}\bra{0} + \ket{0}\bra{-1} + \ket{-1}\bra{0}+ \ket{-1}\bra{-1})\otimes {\bf{I}}/2.
\end{align}

After the SEDOR sequence with free precession time $\tau$, the density matrix is: 

\begin{align}
\rho(0) = \ket{0}\bra{0} +  \\ U_{-1}(\tau/2)\sigma_x U_0(\tau/2)\ket{0}\bra{-1}U_{-1}(\tau/2)^{\dagger} \sigma_x  U_0(\tau/2)^{\dagger} + \\ U_{0}(\tau/2) \sigma_x U_{-1}(\tau/2)\ket{-1}\bra{0}U_0(\tau/2)^{\dagger} \sigma_x U_{-1}(\tau/2)^{\dagger}  + \\ \ket{-1}\bra{-1}.
\end{align}

Where for each term there is an implicit ${\bf{I}}/2$ for the electron spin subspace $\ket{\Uparrow}$ and $\ket{\Downarrow}$. As expected, the population terms stay the same and the coherence terms accumulate phase. Tracing over the electron spins' subspace, our signal is therefore given by $\text{Tr}(S_{x}^{\text{NV}}\rho)$, which is proportional to the trace of the matrix $(U_{-1}(\tau/2) \sigma_{x} U_{0}(\tau/2))^{\dagger} U_{0}(\tau/2) \sigma_{x} U_{-1}(\tau/2)$.

\begin{table*}[!t]
\caption{Frequency components in the SEDOR experiment and their values. The frequency name $\Delta_{i}$ and the form calculated using equation (1) in the main text and the SEDOR pulse sequence are listed as `Frequency' and `Form'. The frequency values in our model, in MHz, from the calculation are listed as `Model (MHz)'. The frequency values extracted from the fit of the time-domain SEDOR data shown in Figure 3(b) are listed as `Fit (3(b)) (MHz)'. Error bars are 95\% CI. For intuition, the states responsible for each frequency component in the SEDOR data are listed as `Relevant states'. Frequency $\Delta_1$ is the dark spin flip-flop rate, corresponding to $\ket{0}_{\text{NV}}$; frequency $\Delta_2$ results from the dipolar magnetic field generated by the dark spins in the $\ket{\uparrow \downarrow}$, $\ket{\downarrow\uparrow}$ subspace, while flip-flops are suppressed by the NV field gradient generated by $\ket{-1}_{\text{NV}}$; frequency $\Delta_3$ is due to the interference of the two paths generated by the NV coherence; and frequency $\Delta_4$ results from the field generated by the dark spins at the NV from the $\ket{\uparrow\uparrow}$, $\ket{\downarrow\downarrow}$ subspace. It is indistinguishable from the component at DC generated by the electron spin dynamics.}
\centering
\begin{tabular}{ |l|l|p{1.2cm}|p{1.5cm}|l|} 
\hline
 Frequency&Form & Model (MHz)&Fit (3(b)) (MHz)
&Relevant states\\ \hline

$\Delta_{1}$&$\sqrt{J_{12}^2 + (\omega_1-\omega_2)^2}$&$0.41(5)$&$0.391(8)$&$\ket{0}_{\text{NV}}\otimes(\ket{\uparrow\downarrow}, \ket{\downarrow\uparrow})$\\

$\Delta_{2}$&$\sqrt{(A_1 - A_2 + \omega_1 - \omega_2)^2 +J_{12}^2}$&$1.85(8)$&$1.790(9)$&$\ket{-1}_{\text{NV}}\otimes(\ket{\uparrow\downarrow}, \ket{\downarrow\uparrow})$\\

$\Delta_{3}$&$\sqrt{(A_1 - A_2 + \omega_1 - \omega_2)^2 +J_{12}^2}-$&$1.44(4)$&$1.430(7)$&$\ket{0, -1}_{\text{NV}} \otimes(\ket{\uparrow\downarrow}, \ket{\downarrow\uparrow})$\\ 

 &$\sqrt{J_{12}^2 + (\omega_1-\omega_2)^2}$& & & \\

$\Delta_{4}$&$A_1 + A_2$&$-0.05(7)$&$0.00(7)$&$\ket{0,-1}_{\text{NV}}\otimes(\ket{\downarrow\downarrow}, \ket{\uparrow\uparrow})$\\ \hline
\end{tabular}
\label{table:S1}
\end{table*}

\begin{table}[h!]
\centering
\begin{tabular}{ |p{2cm}|p{9cm}|p{2cm}| } 
\hline
 Frequency & Amplitude Form & Relative Amplitude\\ \hline
 $\Delta_{1}/2$ & $\frac{1}{\Delta_1^2 \Delta_2^2}\Big(\Delta_2^2(J_{12}^2 - \delta^2) - \Gamma + \Delta_1^2(J_{12}^2-\delta_1^2) + \Delta_1^2 \Delta_2^2\Big) $ &1.00
 
 \\ \hline
 $\Delta_{2}/2$& same as $\Delta_1$ & 1.00
 
 \\
 \hline
 $\Delta_{3}/2$ & $\frac{1}{2\Delta_1^2\Delta_2^2}\Big( \Delta_2^2(\delta^2 - J_{12}^2) + \Gamma + \Delta_1^2 \Delta_2^2 + \Delta_1^2(\delta_1^2 - J_{12}^2) + 4 \delta \delta_1 \Delta_1\Delta_2 \Big)$ & 1.05

 \\ \hline
 $(\Delta_{1}+\Delta_{2})/2$&$\frac{1}{2\Delta_1^2\Delta_2^2}\Big( \Delta_2^2(\delta^2 - J_{12}^2) + \Gamma + \Delta_1^2 \Delta_2^2 + \Delta_1^2(\delta_1^2 - J_{12}^2) - 4 \delta \delta_1 \Delta_1\Delta_2  \Big)$ &0.21
   \\ \hline
 DC&$\frac{1}{\Delta_1^2\Delta_2^2}\Big( \Delta_2^2(\delta^2 - J_{12}^2) + \Gamma + \Delta_1^2 \Delta_2^2 + \Delta_1^2(-\delta_1^2 + J_{12}^2) \Big)$ &1.10
\\ \hline
 
\end{tabular}
\caption{SEDOR frequencies and their relative amplitudes, normalized to the $\Delta_1/2$ frequency component. Here, $\delta \equiv \omega_1 - \omega_2 = 0.14 \text{ MHz}$ is the energy difference between the dark spin states $\ket{\downarrow\uparrow}$ and $\ket{\uparrow\downarrow}$, when the NV population is in $m_s = 0$, $\delta_1 \equiv A_1 - A_2 + \delta = 1.81 \text{ MHz}$ is the energy difference between the dark spin states $\ket{\downarrow\uparrow}$ and $\ket{\uparrow\downarrow}$, when the NV population is in $m_s = -1$, and $\Gamma \equiv \delta^2 \delta_1^2 + \delta^2 J_{12}^2 + \delta_1^2 J_{12}^2 + J_{12}^4$. Note that the $(\Delta_1 + \Delta_2)/2$ component is suppressed (given by the sign of $\delta$). The amplitudes in Figures 3(c) and 4(b) agree with these values within the experimental noise floor.}
\label{table:S2}
\end{table}

The frequency components in this part of the signal are $\Delta_{1}/2 = \frac{1}{2}\sqrt{J_{12}^2 + \delta^2}$, \\$\Delta_{2}/2 = \frac{1}{2}\sqrt{J_{12}^2 + (A_{1} - A_{2} + \delta)^2}$, and $\Delta_{3}/2 = (\Delta_{2} - \Delta_{1})/2$, as expected. There is also a frequency component at DC, in addition to the component at $(A_1 + A_2)/2$ mentioned above. We do not include this component in our analysis, since it is indistinguishable from $(A_1 + A_2)/2$. 

\subsection{Phase modulation}
By sweeping the phase $\phi$ of the last $\pi/2$ pulse, such that $\phi = 2\pi\nu\tau$, any signal component proportional to $\cos{(2\pi(\Delta_{i}\tau/2))}$ will be converted to $\cos{(2\pi(\Delta_{i}/2+\nu)\tau)}+\cos{(2\pi(\Delta_{i}/2-\nu)\tau)}$. These positive and negative frequency components correspond to the polarization of the electron spins. For example, if at the start of the sequence the electrons' spin states begin in the state $\ket{\uparrow\downarrow}$, the NV Bloch vector will rotate clockwise in the transverse plane, with rate 0.84(5) MHz. Conversely, if the electrons' spin states start in the state $\ket{\downarrow\uparrow}$, the NV Bloch vector will rotate counter-clockwise at the same rate. After the last $\pi/2$ pulse, which is responsible for the frequency up-conversion, the corresponding signal frequency is (0.84(5) $ \pm \nu$) MHz, respectively.

\section{Parameter values and error ranges}

The frequency values from the time-domain fit in Figure 3(b) of the main text, as well as the model frequency values, are reported in Table \ref{table:S1}. For $\Delta_{1,2,3}$, the frequency values are extracted from a fit of the absolute value of the FFT of the SEDOR data in Figure 3(b) of the main text, to a sum of three Lorentzian lineshapes. We use a fit in the absolute value of the frequency domain, as opposed to the time domain, in order to inform our $\Delta_{1,2,3}$ frequencies with fewer fit parameters (the phases in the time domain data are sensitive to pulse errors and are therefore free parameters). Although the fits in the frequency and time domain agree within the error ranges [Table \ref{table:S1}], we expect small differences occur because of our approximation of the frequency domain lineshape as Lorentzian. As listed in Table \ref{table:S1} and stated in the main text, we find frequency values for $\Delta_{1,2,3}$ of 0.41(5), 1.85(8), and 1.44(4) MHz, respectively. We use frequencies $\Delta_{1,2,3}$ to extract the parameter values $J_{12}$, $A_1 - A_2$, and $\omega_1 - \omega_2$, as well as the lower bound on $|A_1+A_2|$ of 50 kHz. Due to the existence of multiple solutions to this system of equations, we impose agreement to the splitting in the NV ESR spectrum and the value of $A_1-A_2$ to find our parameter values. The frequency of oscillation in our Hartmann-Hahn experiments [Fig. 4(a) in the main text] confirms our result. Finally, we check for qualitative agreement between the DEER ESR spectrum and a numerical simulation of our model and the DEER ESR pulse sequence. As mentioned in the main text, the DEER ESR lineshape is very sensitive to all parameter values [Fig. \ref{deeresr}]. This allows for finding the value of $A_1 + A_2$ within the lower bound found from the SEDOR experiment and hence $A_1$ and $A_2$. Additionally, the value of $\omega_1$, and therefore $\omega_2 = \omega_1 - 0.14(5) \text{ MHz}$, can be found by centering the central DEER ESR dip with the dip found numerically. We find that $\omega_1$ is indistinguishable from the bare electron Zeeman splitting at $g=2$ (within the experimental error, see below). We note that the exact values of $A_1 + A_2$ and $\omega_1$ are irrelavant to our observation of coherent dynamics, since $A_1 + A_2$ is unresolved in the SEDOR experiment, and the Hamiltonian conserves $|S_z^{(1)} + S_z^{(2)}|$, such that the common-mode Zeeman energy splitting of both spins is inconsequential.

To account for the finite spectral resolution shifting our fit frequency, we add an uncertainty of half of the spectral resolution (approximating this to be the 95\% CI) to the error in the frequency-domain fit. We expect that the small differences between the frequencies extracted from the time domain fit and the frequency domain fit, which are within the experimental uncertainty, are due to approximating the frequency-domain lineshapes as Lorentzians. The error of the $\Delta_4$ frequency is calculated using the estimated error ranges on our parameter values of 50 kHz. The amplitude of a fourth frequency component, $(\Delta_{2} + \Delta_{1})/2$, is suppressed for our parameters, as shown in Table \ref{table:S2}.

According to our model, the frequencies that appear in the SEDOR experiment are various combinations of the parameters added in quadrature, which we use to inform the coupling strengths reported. These measurements have finite widths and degrees of reproducibility, as seen in Figs. 3(b) and 3(c) in the main text. We expect that the variability of our measurements is due to \textit{B}$_0$ field misalignment changing the secular coupling strengths and drive amplitude instability inducing pulse errors during our characterization experiments. In this section, we describe the procedure used to obtain the model parameter error ranges reported in the main text. 

\begin{figure}[h!]
\centering
\includegraphics[scale = 0.5]{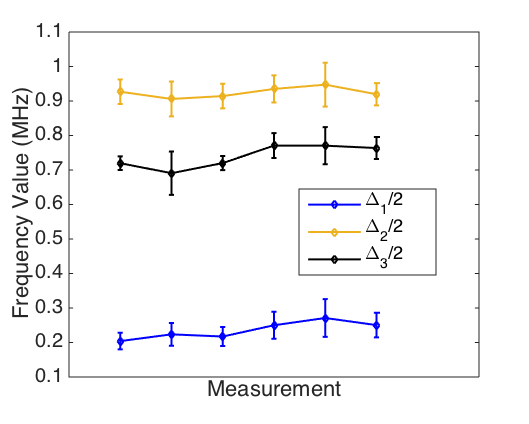}
\caption{\label{fig:many_sedor} SEDOR frequency values $\Delta_{1,2,3}/2$ measured over the data-taking period. Values are extracted from a fit to the FFT of the SEDOR data using a sum of three Lorentzian lineshapes. Error bars account for the 95\% CI of the fit as well as the finite spectral resolution shifting the fit frequency (approximating full-width 95\% CI as the spectral resolution).}
\end{figure}

\begin{figure}[htbp]
\centering
\caption{DEER ESR lineshapes calculated numerically using parameter values within the error range reported in the main text. Black circles are data, and red line is the DEER ESR lineshape calculated numerically for the parameter values given in the main text. Blue lines are DEER ESR results calculated numerically for parameter values listed. (a) Blue line: DEER ESR lineshape calculated numerically, using parameter values listed in the main text, except with $J_{12} = 0.43$ MHz, or 50 kHz more than the $J_{12}$ reported. (b) Blue line: DEER ESR lineshape calculated numerically, using parameter values listed in the main text, except with $J_{12} = 0.33$ MHz, or 50 kHz less than the $J_{12}$ reported. (c) Blue line: DEER ESR lineshape calculated numerically, using parameter values listed in the main text, except with $A_{1} = 0.86$ MHz, or 50 kHz greater than the $A_{1}$ reported. (d) Blue line: DEER ESR lineshape calculated numerically, using parameter values listed in the main text, except with $A_{2} = -0.76$ MHz, or 100 kHz less negative than the $A_{2}$ reported, beyond the error range in our parameters.}{\includegraphics[width =\textwidth]{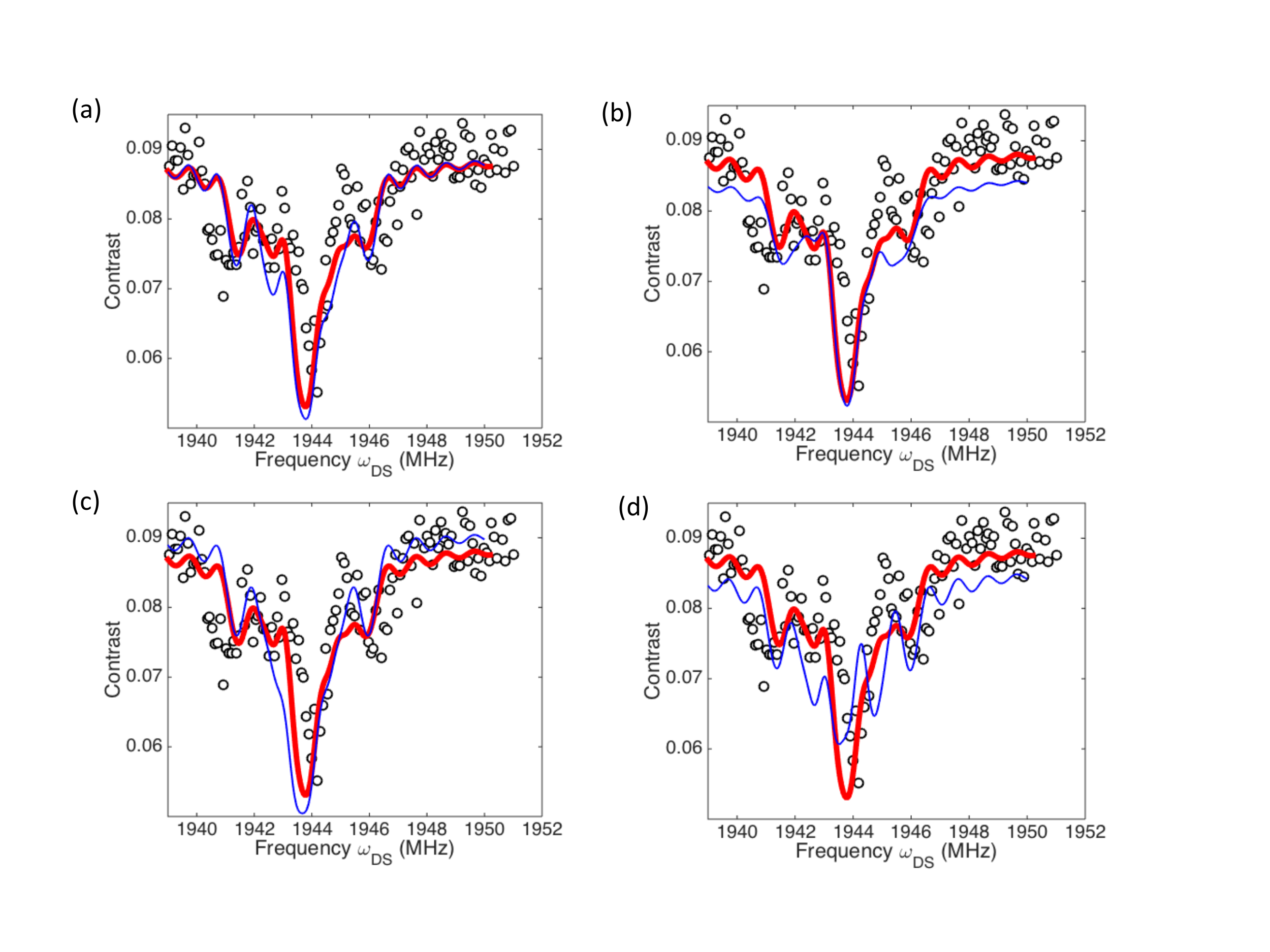}}
\label{deeresr}
\end{figure}

We measure the degree of reproducibility of the SEDOR experiment by repeating the experiment six times over the course of our data-taking period (about 6 months). To extract the SEDOR frequencies $\Delta_{1,2,3}$, we fit the Fourier transform of the SEDOR data to a sum of three Lorentzian curves, and extract the center frequency as well as 95\% CI on the fit (to account for photon shot noise). Due to the finite spectral resolution of our experiment, which is on the order of the widths that we measure, we add a contribution to the error equal to half the frequency spacing in our SEDOR experiments (approximating 95\% CI to be the full frequency spacing). The SEDOR frequencies obtained for the six measurements, with the error ranges found, are plotted in Figure \ref{fig:many_sedor}. For each of the SEDOR frequencies $\Delta_{1,2,3}$, we define the range of the frequency value to be $\text{max}(\Delta_i) + \sigma_{\text{max}(\Delta_i)} - (\text{min}(\Delta_i) + \sigma_{\text{min}(\Delta_i)})$, where $\sigma$ is defined to be the 95\% CI for the relevant measurement. 

Given the form for $\Delta_{1-3}$ reported in the main text, we estimate the drift and error range for each of the parameter values, by taking the average of the ranges of the $\Delta_{1,2}$ values, divided by $2\sqrt{2}$, or approximately 50 kHz.

The error in $\omega_1$, and consequently the common mode error in $\omega_2 = -0.14 \text{ MHz} + \omega_1$ is given by the $B_0$ field misalignment which, as mentioned in Section 1.4 is about 2 degrees. At 694.0 G, this is about 0.5 G or 1.4 MHz. A measurement of our NV's zero field splitting gives a value of 2.7817 $\pm 0.0015$ GHz, adding to the $B_0$ uncertainty. The $\omega_1$ value is about $\gamma_e B_0 +0\substack{+1.5 \\ -2.0} \text{ MHz}$. We find that the DEER ESR lineshape qualitative agreement occurs within this window around the bare electron Zeeman splitting.

\subsection{Comparison of DEER ESR spectra for various coupling strength values}
In this section, we demonstrate the qualitative agreement between our DEER ESR spectrum and our model, within the error range on the parameters reported. As in the main text, we simulate the results of the DEER ESR experiment by numerically calculating the ODMR lineshape under the DEER ESR pulse sequence, for our spin cluster Hamiltonian and parameters. To explore the sensitivity of our lineshape to our parameter values, we repeat the numerical calculation, changing the parameter values one at a time, and compare the results to the same data. Figure \ref{deeresr} demonstrates qualitative agreement between the DEER ESR data and our model for parameter values within the error range reported in the main text. For parameter values simulated beyond the error range, as in Figure \ref{deeresr}(d), the numerically calculated DEER ESR lineshape clearly disagrees with the DEER ESR data obtained. Changing the relative detuning $\omega_1 - \omega_2$ between the electron spins and the other coupling strength $A_{1}$, at and beyond the error bar ranges, gives similar qualitative results.

\section{SEDOR frequency component at the phase cycling frequency due to pulse errors}

In this section we characterize the amplitude of the NV-electron spin SEDOR signal at the phase modulation frequency, $\nu$, when there are imperfect pulses. Although there is a component of the SEDOR signal predicted to be at $\nu$, the amplitude can depend on pulse errors, as calculated below. The SEDOR pulse sequence is illustrated in Figure 3(a) in the main text. The intuitive idea is that if, due to detuned driving, the electron spin is not fully flipped, then there is some component of NV coherence that is left oscillating purely at the phase modulation frequency. Here we calculate the amplitude of that component as a function of the Rabi frequency and detuning. 

In this section we evaluate the density matrix at various points in the evolution of the sequence, instead of multiplying unitaries and taking a trace at the end, in order to account for the finite electron spin coherence time. 

\subsection{Hamiltonian and pulse sequence}

To gain a qualitative understanding of the component at $\nu$ due to pulse errors, we imagine an NV (NV) and single electronic dark spin (DS) system. We assume that their resonance frequencies are detuned from each other much more than both their linewidths and coupling strength $A_{||}$. Additionally, we incorporate an RF drive on the NV and dark spin, and we assume that the drive on the NV is resonant (we are ignoring NV hyperfine splitting), and that the dark spin drive is detuned by $\delta$. Treating the NV in a 2x2 subspace, we absorb the extra $A_{||}/2 S_{z}^{\text{DS}}$ term (the third term in equation (1) in the main text) into $\delta$:

\begin{equation} 
H = A_{||} S_{z}^{\text{NV}} S_{z}^{\text{DS}} + H_{RF}(t).
\end{equation}

The pulse sequence is a Hahn echo on the NV, with the NV $\pi$ pulse coincident with a dark spin detuned $\pi$ pulse. The full free precession time is 2t. We assume perfect pulses on the NV, and a $\pi$ pulse of duration T on the dark spin. The goal is to see if there is any component of the NV coherence at the end of the pulse sequence that does not oscillate, i.e., a component that will be upconverted to $\nu$ after adding phase modulation.

\subsection{Dynamics} 

We start with a polarized state on the NV, $S_{z}^{\text{NV}} + {\bf{I}}^{\text{NV}}/2$ and a fully mixed state on the dark spin, ${\bf{I}}^{\text{DS}}/2$. After the first $\pi/2$ pulse on the NV, along y, the density matrix is then

\begin{equation}
\rho = (S_{x}^{\text{NV}}+{\bf{I}}^{\text{NV}}/2))\otimes {\bf{I}}/2.
\end{equation}

(We use the notation $S_{x} \equiv S_{x}\otimes {\bf{I}}/2$ here for simplicity.) 

After the first part of the free evolution time we have: 
\begin{equation}
\rho(t) = S_{x}^{\text{NV}} \cos{A_{||}t/2} + 2 S_{y}^{\text{NV}}S_{z}^{\text{DS}} \sin{A_{||}t/2}.
\end{equation}

Next we have a $\pi$ pulse on both the NV (about x) and the dark spin (about x), with the dark spin pulse detuned by $\delta$ and for duration $T << 1/A_{||}$, such that $H_{eff}^{\text{DS}} = \delta S_{z}^{\text{DS}} + \Omega_{\text{DS}}S_{x}^{\text{DS}}$. We assume also that the difference in the NV and dark spin resonance frequencies (order 1 GHz) is much greater than either Rabi frequency. 

\begin{align}
\rho(t + T) = S_{x}^{\text{NV}}\cos{A_{||}t/2} 
\\- 2S_{y}^{\text{NV}}\sin{A_{||}t/2}\Big(e^{-i(\delta S_{z}^{\text{DS}} + \Omega_{\text{DS}} S_{x}^{\text{DS}})T}\Big) S_{z}^{\text{DS}} \Big(e^{i(\delta S_{z}^{\text{DS}} + \Omega_{\text{DS}}S_{x}^{\text{DS}})T}\Big)
\end{align}

To solve for the dark spin dynamics under the detuned pulse we need to go into a tilted frame to diagonalize the dark spin Hamiltonian. Choosing $U = e^{-i\theta S_{y}^{\text{DS}}}$ we have $H_{eff}^{\text{DS}} = \bar\Omega_{\text{DS}} S_{z}^{\text{DS}}$ where $\bar \Omega_{\text{DS}} = \sqrt{\Omega_{\text{DS}}^2 + \delta^2}$ and $\theta = \arctan{-\Omega_{\text{DS}}/\delta}$. After going into the tilted frame and applying the detuned pulse, we account for evolution during the second half of the free precession time. After transforming back into the un-tilted frame, we drop all $S_{x,y}^{\text{DS}}$ terms, since the dark spin $T_{2}^* << 1/A_{||}$. Allowing for the second half of the free precession time, we find a final density matrix of:

\begin{align}
    \rho(2t + T) = S_{x}^{\text{NV}} \cos^2{A_{||}t/2} + 2 S_{y}^{\text{NV}} S_{z}^{\text{DS}} \cos{A_{||}t/2} \sin{A_{||}t/2}
    \\- \cos^2{\theta}\sin{A_{||}t/2}\Big(2S_{y}^{\text{NV}}S_{z}^{\text{DS}}\cos{A_{||}t/2} - S_{x}^{\text{NV}} \sin{A_{||}t/2}\Big)
    \\- \cos{\bar\Omega_{\text{DS}} T} \sin^2{\theta} \sin{A_{||}t/2} \Big(2 S_{y}^{\text{NV}} S_{z}^{\text{DS}} \cos{A_{||}t/2} - S_{x}^{\text{NV}} \sin{A_{||}t/2}\Big)
\end{align}

Since the dark spin is unpolarized, any term that evolves as $S_{z}^{\text{DS}}$ will add a mixed state contribution and therefore not contribute to any NV coherence oscillation. So the ``signal'' terms are:

\begin{align}
    \rho(2t + T) \sim S_{x} (\cos^2{A_{||}t/2} + \cos^2{\theta}\sin^2{A_{||}t/2} + \cos{\bar\Omega T} \sin^2{\theta} \sin^2{A_{||}t/2}).
\end{align}

As a check, if we have $\bar \Omega T = \pi$ and $\delta = 0$ meaning $\theta = \pi/2$, we would find $\rho(2t + T) \sim S_{x}(\cos^2{A_{||}t/2} - \sin^2{A_{||}t/2}) = S_{x}(\cos{A_{||}t})$ as desired. 
\subsection{Results}

For $\delta \neq 0$, we set $\bar \Omega_{\text{DS}} T = \pi$ and calculate the resulting NV coherence: 

\begin{align}
    \rho(2t + T) \sim S_{x} (\cos^2{A_{||}t/2} + \sin^2{A_{||}t/2}(\cos^2{\theta} - \sin^2{\theta})
    \\ = S_{x}(\cos^2{\theta} + \sin^2{\theta}\cos{A_{||}t})
\end{align}

Clearly there is a component that does not oscillate, i.e., the $\cos^2{\theta}$ component. After applying the last $\pi/2$ pulse and phase modulation, this component will oscillate purely at the phase modulation frequency. This component at DC has an amplitude of $\cos^2{\theta} = \delta^2/(\Omega_{\text{DS}}^2 + \delta^2)$. If $\bar\Omega_{\text{DS}} T = (2n+1)\pi/2$, we are left with: 

\begin{align}
\rho(2t + T) \sim S_{x}(\cos^2{A_{||}t/2} + \cos^2{\theta}\sin^2{A_{||}t/2})
\\ = S_{x}(\cos^2{\theta}+(1/2)\sin^2{\theta} - \sin^2{\theta}\cos{A_{||}t})
\end{align}

In this case we find a component proportional to $(\delta^2 + (1/2)\Omega_{\text{DS}}^2)/(\delta^2 + \Omega_{\text{DS}}^2)$ which is always significant.

As desired, the only case where most of the population is actually oscillating at the dipole coupling strength $A_{||}$ is when $\delta/\Omega_{\text{DS}} << 1$ and $\bar \Omega_{\text{DS}} T = n\pi$ with $n$ odd. 

In our experiment, $\delta$ is of order 1 MHz and $\Omega$ is of order 13 MHz. For a $\pi$ pulse, the extra amplitude component in $\nu$ is approximately $\cos^2{\theta}$, which for these parameters is about 0.6$\%$.

\section{S quantum number of the electron spins}

The electron spins are determined to be S = 1/2 by comparing the measured NV Rabi frequency, $\Omega_{\text{NV}}$, to the electron spin Rabi frequency, $\Omega_{\text{DS}}$, for known resonant drive field amplitudes. The NV spin transitions are properly treated in a 2x2 subspace, although the NV electronic spin has S = 1, because the other NV spin transition is far off resonance for our experimental conditions. Thus, the normalized magnitude of the NV $S_{x, y}$ matrix elements are $1/\sqrt{2}$. For S = 1/2, the normalized magnitude of the $S_{x,y}$ matrix elements are $1/2$. From our experiments we find $\Omega_{\text{NV}}/\Omega_{\text{DS}} = \sqrt{2}$, as expected for S = 1/2 electron spins. This technique for spin quantum number identification is commonly used in EPR \cite{schweiger}.

We are careful to use the same electronics and carrier frequency for both measurements, by tuning the $B_{0}$ field between measurements such that the resonance frequency of the NV $m_s = 0 \rightarrow m_s = -1$ transition during the $\Omega_{\text{NV}}$ measurement is equal to the electron spin resonance frequency during the $\Omega_{\text{DS}}$ measurement. In this case, the resonance frequency of both spins was 927.2 MHz. 

\begin{figure}[h!]
\centering
\includegraphics[scale = 0.3]{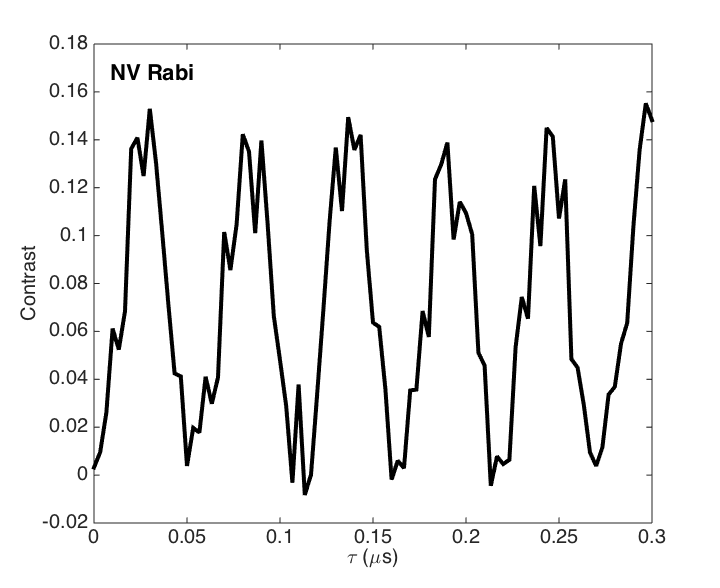}
\caption{Rabi flopping of the NV, at a resonance frequency of 927.2 MHz. Here, $\tau$ is the NV pulse duration, the Y axis is contrast. A sinusoidal fit to the data gives a Rabi frequency of 18.5(2) MHz.}
\label{NVrabi}
\end{figure}

Running a Rabi experiment on the NV, we see coherent oscillations [Fig. \ref{NVrabi}]. The Rabi frequency extracted from fitting the data is 18.5(2) MHz. 

\begin{figure}[h!]
\centering
\includegraphics[scale = 0.3]{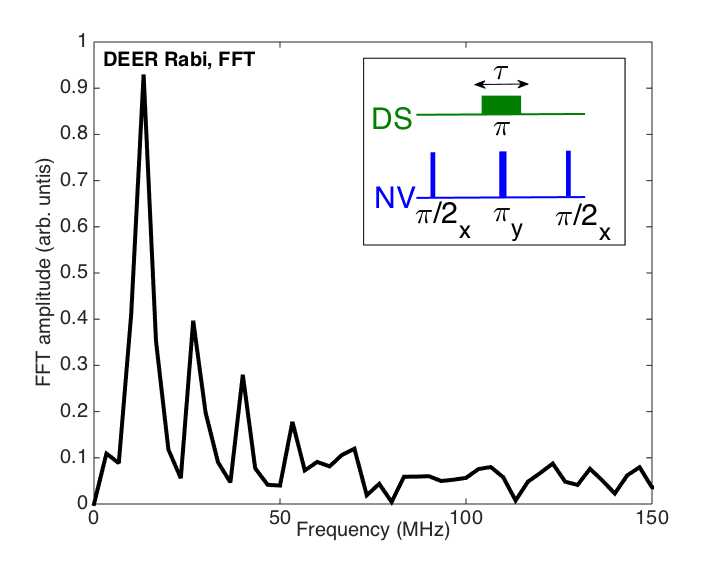}
\caption{FFT of the DEER Rabi experiment on the electron spins, at a spin resonance frequency of 927.2 MHz. The primary frequency of 13.3 MHz corresponds to the electron spin Rabi frequency, and the harmonics are from pulse errors in the sequence due to the long NV-electron spin coupling time.}
\label{deerrabi}
\end{figure}

We immediately perform a DEER Rabi pulse sequence to extract the Rabi frequency of the electron spins, with the field $B_0$ set such that the resonance frequency of the electron spins is at about 927 MHz, to avoid frequency-dependent power delivery of the setup affecting our S value measurement. We see multiple frequency components in the resulting FFT [Fig. \ref{deerrabi}] of the time-domain data, due to the fact that the coupling time $\tau$ is longer than the interaction period $1/A_{1} \approx 1/A_{2}$. We find a primary frequency of 13.3(1) MHz, with the harmonics from this feature appearing in the spectrum. Since $\Omega_{\text{NV}}/\Omega_{\text{DS}} = \sqrt{2}$ within their error bars, we conclude that both electron spins are S = 1/2. 

These measurements show that the dark spins are S = 1/2 electronic spins with no nuclear spins present.  To our knowledge, the only possibility in diamond for stable S = 1/2 electronic defects with no nuclear spins is the V+ defect. 

\section{Zeeman spectroscopy of the electron spins}

\begin{figure}[h!]
\centering
\includegraphics[scale = 0.7]{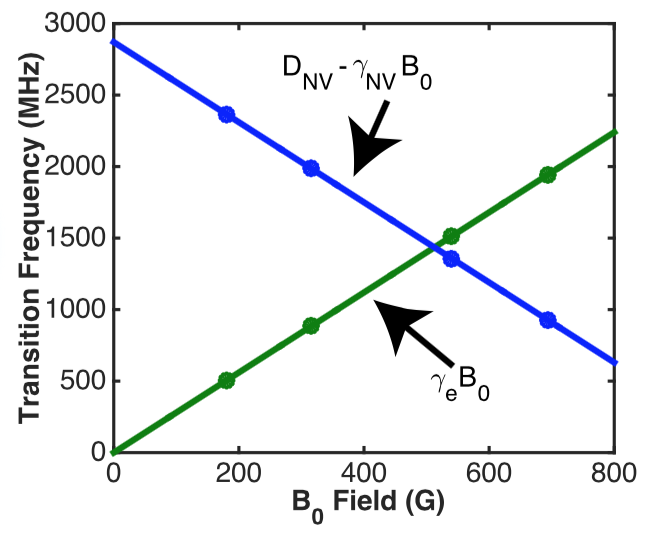}
\caption{\label{fig:figS3} NV and electron spin resonance frequencies for various $B_0$ field values. Blue and green dots are NV ESR and electron spin ESR data, and blue and green lines are theoretical predictions describing the NV transition frequency (neglecting hyperfine and electron spin coupling), and the bare electron spin transition frequency (neglecting NV and electron spin coupling). The NV and electron spin splittings of order 1 MHz are much smaller than the size of the dots.}
\end{figure}

The electron spins' resonance frequencies closely follow the bare electron $g=2$ value in our experiment. In general, the order of magnitude $\sim 10^{-4}$ shift from $g = 2$ that we find is consistent with other g anisotropy values for other S = 1/2 defects in diamond \cite{newton paper 1,newton paper 2,baker paper}. To demonstrate their bare electron character, we perform Zeeman spectroscopy by changing the value of the $B_0$ field and measuring the transition frequencies. We move a permanent magnet nearby and align it to the NV axis to within a couple degrees [Section 1.4]. We then measure the strength of the magnetic field via the NV $m_s = 0 \rightarrow m_s = -1$ transition frequency, and perform DEER ESR experiments to find the transition frequencies of the electron spins. Plotting the NV transition frequency as well as the central dip of the DEER ESR spectrum and with the $B_0$ field value shows definitive $g=2$ character [see Fig. \ref{fig:figS3}]. 

\section{Contrast definition}
In order to mitigate noise from laser power drifts over our measurements, we symmeterize every pulse sequence reported in the main text (except for the NV ESR, for which there is no corresponding measurement). For a given sequence that ends in a backprojection on the NV of $\pi/2_{\phi}$, we repolarize the NV to $m_s = 0$, and repeat the sequence, ending with a pulse of $\pi/2_{\phi + \pi}$ on the NV. For example, a spin echo sequence consists of: $\pi/2_{x} - \pi_{x} - \pi/2_{x} \text{ [532 nm laser] } \pi/2_{x} - \pi_{x} - \pi/2_{-x} \text{ [532 nm laser] }$. We read out the signal at each laser pulse. If the amount of photons acquired at the laser pulses are $Y_1$ and $Y_2$, respectively, then our contrast is defined to be $(Y_2 - Y_1)/(Y_2 + Y_1)$. This scheme allows us to retain sensitivity while subtracting common-mode noise. In our numerical simulations, shown in Figs. 2(b) and 4(a) in the main text, we allow the contrast to be a free parameter and find the best qualitative fit to the amplitude of the features. Background fluorescence from the buildup of dust on our sample, as well as laser power drifts affecting the optimal readout pulse duration, can change the contrast on our experimental timescales of days. Nonetheless, we find reasonable agreement between the relative size of the measured features and the size of the signals from our numerical simulations.

\section{Measurements at misaligned fields}

We attempted to recover the dark spin spatial distance information using a technique found in \cite{alex single proton}, wherein the direction of the bias field (at any magnitude, i.e. away from the ESLAC or GSLAC) changes the secular dipolar coupling strengths as the NV and dark spin quantization axis become misaligned. The authors found that for a 15 degree misalignment away from the NV axis, only one frequency appeared in the SEDOR experiment, at around 1 MHz: 

\begin{figure}[h!]
\centering
\includegraphics[scale = 0.4]{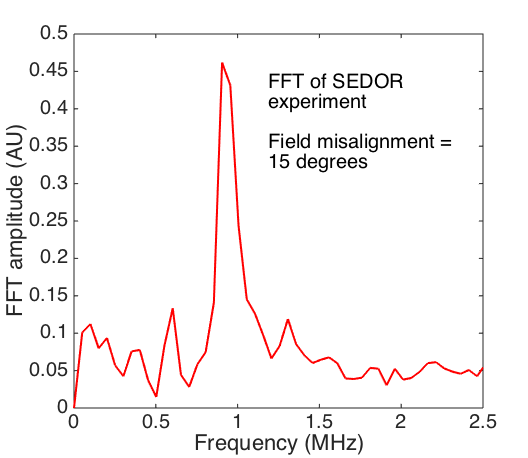}
\caption{\label{fig:first_misalign} FFT of the SEDOR experimental results, under a field misalignment of 15 degrees from the NV axis. Only one frequency appears, implying that the electron spin-spin flip-flops are suppressed.}
\end{figure}

This observation implies that the NV-dark spin coupling magnitudes remained indistinguishable, and that dark spin flip-flops became suppressed, due to either a reduction in $J_{12}$, an increase in $\omega_1 - \omega_2$ from g anisotropy, or both. To attempt to recover the electron spin-spin flip-flops at another angle, we repeated the SEDOR experiment, with the misalignment from the NV axis  again at 15 degrees, but the azimuthal angle $\phi$ changed by 7 degrees from the previous measurement. The SEDOR FFT once again shows one peak: 

\begin{figure}[h!]
\centering
\includegraphics[scale = 0.4]{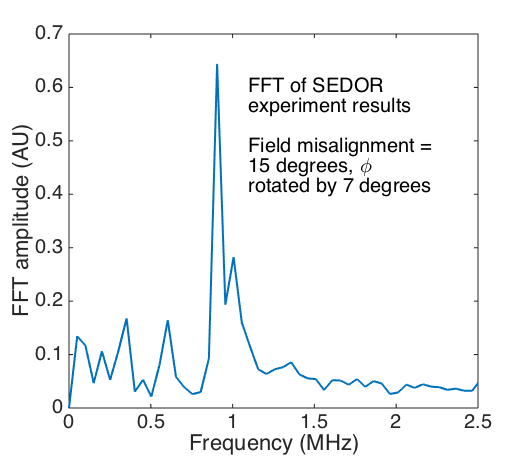}
\caption{\label{fig:second_misalign} FFT of the SEDOR experimental results, under a field misalignment of 15 degrees from the NV axis. Now the azimuthal angle $\phi$ is changed by 7 degrees. Only one frequency appears, implying that the electron spin-spin flip-flops are suppressed.}
\end{figure}

The electron spin-spin flip-flops remained suppressed, and the Hamiltonian cannot be uniquely identified. Quantifying this suppression effect, by estimating the g anisotropy and $J_{12}$ at very small misalignment angles, would require field alignment accuracy and precision better than the 2 degree precision reported here.

\end{document}